\documentclass[showpacs,preprintnumbers,amsmath,amssymb]{revtex4}

\usepackage{graphicx}
\usepackage{dcolumn}
\usepackage{bm}
\usepackage{url}
\usepackage{epsfig}
\usepackage{multirow}
\usepackage{tensor}
\usepackage{empheq}
\usepackage{amsmath}
\usepackage{color}
\definecolor{myblue}{rgb}{.8, .8, 1}
\usepackage{makecell,diagbox}

\def\be{\begin{equation}}
\def\ee{\end{equation}}
\def\ba{\begin{eqnarray}}
\def\ea{\end{eqnarray}}

\newcommand{\fr}[2]{\frac{#1}{#2}}

\newcommand{\PRD}{Phys.\ Rev.\ D}

\newcommand{\MNRAS}{Mon.\ Not.\ Roy.\ Astron.\ Soc.}

\def\ga{\mathrel{\raise.3ex\hbox{$>$\kern-.75em\lower1ex\hbox{$\sim$}}}}
\def\la{\mathrel{\raise.3ex\hbox{$<$\kern-.75em\lower1ex\hbox{$\sim$}}}}

\begin{document}


\leftline{KIAS-P15056}

\title{Cross correlation of Cosmic Microwave background and Weak Lensing}


\author{Seokcheon Lee}

\affiliation{School of Physics, Korea Institute for Advanced Study, Heogiro 85, Seoul 130-722, Korea}



\begin{abstract}
The integrated Sachs-Wolfe (ISW) effect and its non-linear extension Rees-Sciama (RS) effect provide us the information of the time evolution of gravitational potential. The cross-correlation between the cosmic microwave background (CMB) and the large scale structure (LSS) is known as a promising way to extract the ISW (RS) effect. It is known that the RS effect shows the unique behavior by changing the anti-correlated cross correlation between the CMB and the mass tracer into the positively correlated cross correlation compared to the linear ISW effect. We show that the dependence of this flipping scale of the cross-correlation between RS and weak lensing on dark energy models. However, there exists the degeneracy between DE and $\Omega_{\rm{m}0}$ which might be broken by redshift dependent observables. The cross-correlation between the momentum field and the density field might be served as the better observable to be used for this purpose.  

\end{abstract}



\maketitle

\section{Introduction}

After the recombination, photons free-stream, but they are still affected by gravitational redshifts. In the dark energy dominated universe, if a photon falls into a potential well along its way to the observer, it must climb out of a shallower potential well than it fell into. This contribution to the CMB anisotropy from redshifts in the time-varying gravitational potential wells is called the integrated Sachs-Wolfe (ISW) effect \cite{ISW}. Non-linear evolution of density fluctuations also contribute to the ISW anisotropy and it is called the Rees-Sciama (RS) effect \cite{RS}. 
Thus, the late time ISW (RS) effect is sensitive to dark energy and its detection presents an independent signature of dark energy. The cross-correlation between the ISW (RS) and the local matter density (or its tracer) is known to be useful to isolate the ISW (RS) effect from the CMB \cite{9510072}. From now on, we called RS effect including ISW effect. If one can remove the any astrophysical foreground from the CMB measurement,  then the cosmic microwave background temperature anisotropies are produced by various sources including the deflection of CMB photons by the gravitational lensing, the thermal and the kinetic Sunyaev-Zel$^{'}$dovich (t/kSZ) effects,  and the RS effect, in addition to its primordial temperature anisotropies on small angular scales. Thus, the cross-correlation of CMB temperature anisotropies, $\Theta \equiv \Delta T / T $ with the density tracer of the large scale structure (LSS), $\delta_{\rm{tr}}$ can be written as \cite{14045102} 
\be \Bigl \langle \Theta \, \delta_{\rm{tr}} \Bigr \rangle = \Bigl \langle \Bigl(\Theta_{\rm{pri}} + \Theta_{\rm{lens}} + \Theta_{\rm{t(k)SZ}} + \Theta_{\rm{RS}} \Bigr)  \, \delta_{\rm{tr}} \Bigr \rangle \, . \label{crocor} \ee
Among these, one can ignore the primordial anisotropies of CMB, $\Theta_{\rm{pri}}$ in the cross correlation with the tracer of the LSS because their generations have very different epochs. In principle, one can also remove tSZ, $\Theta_{\rm{tSZ}}$  from the temperature anisotropies using the multi-frequency observation because of its frequency dependence on the amount of intensity distortion of the photon \cite{9808050}. Finally, one needs to consider the correlation between lensed CMB, kSZ, RS, and density tracer. Even though the kSZ and the mass distribution are not correlated in the linear regime because peculiar velocities of clusters are random in their directions along the line-of-sight, the non-linear cross-correlation between the kSZ and the LSS is correlated. However, it is two orders of magnitudes smaller than that of the RS and the LSS \cite{0309337}.   

Thus, RS effect can be detected by cross-correlating with some tracers of the large scale structure. As density tracers, one can use either the galaxy distribution \cite{9510072, 9704043, 08094488, 10030974, 10101096, 12107513} or the weak lensing convergence, $\kappa$ \cite{07111696, 09062317, 12043789}. There have been observations to detect the cross-correlation between the RS effect and the LSS \cite{9610160, 0111281, 0305001, 0307249, 0308260, 0401166, 08014380, 08121025, 10102192, 12082350, 12092125, 13035079, 14075623, 150201595}. 

The angular power spectrum between the RS effect and the weak lensing convergence can be used to investigate the cross-correlation between two quantities \cite{07111696}. In order to calculate the non-linear matter power spectrum, one can use the N-body simulation, the halo model \cite{0206508}, or the standard perturbation theory (SPT) \cite{0112551}.  We use the third order SPT to calculate the non-linear power spectrum, as it provides the exact calculation in the quasi-linear regime. This method can be easily extended to various cosmological models compared to the N-body simulation and the halo model. Also, the exact solutions for SPT of various dark energy models is known \cite{14077325} and one can reinforce the previous calculations for the cross-correlation. In this exact solutions, one can obtain the fully consistent third-order density fluctuation and kernels for the general dark energy models without using the Einstein-de Sitter universe assumption. Thus, this result is robust for any dark energy model and can be used for the analytic prediction for the cross-correlation. Compared to the cross-correlation of the linear ISW effect and the LSS, the cross-correlation of the RS effect and the LSS has the unique feature of changing from the anti-correlation to the positive correlation. Previously, we show the dark energy dependence on the location of this flipping scale by using the exact 3rd-order SPT and claimed this scale might be used as a standard ruler to investigate the dark energy model \cite{150703725}. 

In this paper, we reinvestigate the cross-correlation of the RS effect and the weak lensing convergence with the exact 3rd-order SPT. In the next section, we briefly review the RS-$\kappa$ angular power spectrum. In the section 3, we obtain angular power spectra for the different dark energy models and show the dark energy dependence on the flipping scales. We conclude in section 4.

\section{Cross-Correlation of the RS effect and the WL}

In this section, we briefly review how to obtain the angular power spectrum of the RS and the weak lensing convergence, $\kappa$.
The CMB temperature anisotropies due to the RS effect in the flat Universe is given by the line of sight integral of the change in the gravitational potential to the last scattering surface,
\be \Theta^{\rm{RS}}(\hat{n}) = \sum_{l,m} a_{lm}^{\rm{RS}} Y_{lm}(\theta,\phi) = \fr{1}{c} \int_{\eta_{\rm{ls}}}^{\eta_{0}} d \eta e^{-\tau} (\Phi' + \Psi') [k, \eta, \hat{n}(\eta_0-\eta)]   \simeq \fr{2}{c} \int_{\eta_{\rm{ls}}}^{\eta_{0}} d \eta \Phi' [k, \eta, \hat{n}(\eta_0-\eta)]\label{ThetaISW} \, , \ee
where $\eta$ is the conformal time, $\eta_0$ being today, $\eta_{\rm{ls}}$ being recombination, $\tau$ is the optical depth, primes mean the derivatives with respect to the conformal time, $\Phi$ is the Newtonian potential, and $\Psi$ is the spatial curvature perturbation, respectively. We restrict our consideration to the perfect fluid with the general relativity under the instantaneous reionization assumption. Thus, we ignore the anisotropic stress tensor and the optical depth in the last equality of the above equation (\ref{ThetaISW}). From this one obtains the spherical harmonic coefficients for the RS component of CMB
\be a_{lm}^{\rm RS}(\hat{n}_0,\eta_0) = \fr{8\pi}{c} i^l \int \fr{d^3 k}{(2\pi)^3} Y_{lm}^{\ast} (\Omega_{\hat{k}}) \int_{\eta_{0}}^{\eta_{\rm{ls}}} d \eta \Phi' j_{l}(k c (\eta_0 - \eta)) \label{almRS} \, ,\ee
where $j_{l}$ is a spherical Bessel function. The Newtonian potential $\Phi$ and its time derivative are given by the Poisson equation
\ba \Phi(\vec{k},\eta) &\simeq& -\fr{3}{2} \fr{\Omega_{\rm m0}}{a} \Bigl(\fr{H_0}{c k} \Bigr)^2 \delta (\vec{k},\eta) \label{Phi} \, , \\ 
\Phi'(\vec{k},\eta) &\simeq& -\fr{3}{2} \fr{\Omega_{\rm m0}}{a} \Bigl(\fr{H_0}{c k} \Bigr)^2 \Bigl(\delta' (\vec{k},\eta) - a H \delta (\vec{k},\eta) \Bigr)\label{Phip} \, ,\ea
where $\Omega_{\rm m0}$ is the present matter energy contrast, $H_0$ is the present value of the Hubble parameter, and $\delta$ is the matter density fluctuation, respectively.

The distribution of total matter is probed by gravitational lensing observations. From the Poisson's equation, the amplitude of distortion in galaxy images due to WL so-called the convergence, $\kappa$ in any particular direction on the sky $\hat{n}$ at the specific redshift, $z_s$ is equal to the projected mass in the Born limit,
\be \kappa (\hat{n}, z_s)  = \int_{0}^{r_{s}} W(r) \delta(r) dr \,\, , {\rm where} \,\, W(r) = \fr{3}{2} \Omega_{m} H_0^2 \int_{0}^{r_{s}} dr \fr{r(r_s - r)}{r_s} \fr{1}{a} \label{kappazs} \, . \ee
In order to take into account a distribution of sources at various redshifts contributing to the weak gravitational lensing (WL), one needs to convolve the convergence, $\kappa(z_s)$ with a normalized radial source distribution function, $n(z_s)$,
\be \kappa (\hat{n}) = \int_{0}^{z_s} dz_s n(z_s) \kappa (\hat{n}, z_s) \label{kappa} \, . \ee
This distribution of source galaxies often follows a bell-shape with the peak redshift depends on the depth of the survey. We adopt the parametrization of the radial distribution of source galaxies as \cite{Efstathiou, 0003014}, 
\be n(z) = A z^2 \exp[-(z/z_0)^{\beta}] \label{nz} \, , \ee
where the normalization factor, $A$, is determined by $\int_{0}^{\infty} n(z) dz = 1$. There also exist more general forms of $n(z)$ \cite{0404349, 0511090}.  We adopt $\beta$ and $z_0$ values as in reference \cite{07111696}.  
Again, one can obtain the spherical harmonic coefficients for the convergence
\be a_{lm}^{\kappa} = 4\pi i^l \int \fr{d^3 k}{(2\pi)^3} k^2 Y_{lm}^{\ast}(\Omega_{\hat{k}}) \int_{0}^{r_{s}} dr \fr{r(r_s - r)}{r_s} \Phi_{k} (r) j_{l}(kr) \label{almkappa} \, , \ee
where we use the Poisson's equation. 

Then, one can obtain the cross-correlation power spectrum between RS and convergence, $\kappa$
\ba C_{l}^{\rm{RS}-\kappa}(z_s) &=& \Bigl \langle a_{lm}^{\rm{RS}}(\vec{k}) a_{l'm'}^{\kappa \ast}(\vec{k}',z_s) \Bigr \rangle = \fr{4}{\pi} \int dk k^4 \int_{0}^{r_{\ast}} dr \int_{0}^{r_{s}} dr' \fr{r'(r_s - r')}{r_s}  P_{\Phi\Phi'}(k,r,r') j_{l}(kr) j_{l}(kr')  \nonumber \\
&\simeq& 2l^2  \int_{0}^{r_{s}} dr' \fr{r_s - r}{r^3 r_s}  P_{\Phi\Phi'}(k= \fr{l}{r}, r) \Bigl |_{k=l/r} \label{ClRSkappazs} \, , \ea
where the Limber's approximation is used in the last equality and define 
\be \Bigl \langle \Phi_{\vec{k}}(r) \Phi_{\vec{k}'}'(r') \Bigr \rangle = (2\pi)^3 P_{\Phi \Phi'}(k,r,r') \delta_{\rm{D}}(\vec{k} - \vec{k}') \label{PPhipPhi} \, ,\ee where $\delta_{\rm{D}}$ is a Dirac delta function. One can obtain the total cross-correlation, $C_{l}^{\rm{RS}-\kappa}$, by integrating $C_{l}^{\rm{RS}-\kappa}(z_s)$ with a weight of source distribution $n(z_s)$ for a given weak-lensing survey
\be C_{l}^{\rm{RS}-\kappa} = \int_{0}^{z_s} dz_s n(z_s) C_{l}^{\rm{RS}-\kappa}(z_s) \label{Cl} \, . \ee

\section{Angular Power Spectrum of the RS effect and the WL}

In order to obtain the final result of the cross-correlation of the RS effect and $\kappa$ in Eq.(\ref{Cl}), one needs to calculate the power spectrum of $P_{\Phi \Phi'}$ in Eq.(\ref{PPhipPhi}). From Eqs. (\ref{Phi}) and (\ref{Phip}), one obtains
\be P_{\Phi \Phi'} = \fr{9}{4} \Bigl(\fr{\Omega_{\rm m0}}{a} \Bigr)^2 \Bigl(\fr{H_0}{c k} \Bigr)^4 \Bigl( P_{\delta \delta'} - a H P_{\delta\delta} \Bigr) \label{PPhipPhi2} \, . \ee
If one uses the standard perturbation theory up to the 3rd order, then $P_{\delta \delta'}$ and $P_{\delta\delta}$ are given by \cite{150703725}
\ba P_{\delta \delta}(k,\eta) &=& D_{1}^2(\eta) P_{11}(k) + 2 D_{1}^4(\eta) \int d^3 q P_{11}(q) \Biggl[ P_{11}(|\vec{k}-\vec{q}|) \Bigl[ F_{2}^{(s)}(\vec{q}, \vec{k}-\vec{q}, \eta) \Bigr]^2 \nonumber \\ &&+ 3 P_{11}(k) F_{3}^{(s)} (\vec{q},-\vec{q},\vec{k},\eta) \Biggr] \label{Pdeltadelta} \, , \\
P_{\delta \delta'}(k,\eta) &=&  - D_{1}(\eta) D_{1}^{'}(\eta)  P_{11}(k) - D_{1}^3(\eta) D_{1}^{'}(\eta)  \Biggl[ P_{11}(k) \int d^3 q P_{11}(q) \Bigl[ 3 F_{3}^{(s)} (\vec{q},-\vec{q},\vec{k},\eta) + 3 G_{3}^{(s)}(\vec{q},-\vec{q},\vec{k},\eta) \Bigr]  \nonumber \\ 
&&+ 2 \int d^3 q P_{11}(q) P_{11}(|\vec{k}-\vec{q}|) F_{2}^{(s)} (\vec{q},\vec{k}-\vec{q},\eta) G_{2}^{(s)} (\vec{q},\vec{k}-\vec{q},\eta) \nonumber \\ 
&& + 2 \int d^3 q \Bigl[ F_{2}^{(s)} (\vec{q},\vec{k}-\vec{q},\eta) P_{11}(q) P_{11}(|\vec{k}-\vec{q}|)  + G_{2}^{(s)} (\vec{k},\vec{k}-\vec{q},\eta) P_{11}(k) P_{11}(|\vec{k}-\vec{q}|) \nonumber \\ &&+ F_{2}^{(s)} (\vec{k},-\vec{q},\eta) P_{11}(q)P_{11}(k) \Bigr] \alpha(\vec{q},\vec{k}-\vec{q}) \Biggr]\label{Pdotdeltadelta} \, , \ea
where $D_{1}$ is the growth factor, $F^{(s)}_{2,3} \, (G^{(s)}_{2, 3})$ are the exact symmetric 2nd and 3rd order kernels of the density fluctuation (the divergence of the peculiar velocity), and $\alpha(\vec{k}_1,\vec{k}_2) = \fr{\vec{k}_{12}\cdot\vec{k}_1}{k_1^2}$. One can obtain the exact values of them numerically for any model \cite{14075623}. Due to the non-linear effect, the cross-correlation of $\Phi$ and $\Phi'$ changes from the anti-correlated to the positively-correlated. This flipping scale depends on the dark energy model \cite{150703725}. We obtain the linear power spectrum $P_{11}$ from CAMB \cite{camb} with $(\Omega_{b} h^2, \Omega_{c}h^2, n_{s}, A_{s}, h) =$ (0.0226, 0.1244, 0.96, 2.1 $\times 10^{-9}$, 0.7). This corresponds to $\Omega_{m0} = 0.3$ and $\Omega_{\rm{DE}} = 0.7$. 

\begin{figure}
\centering
\vspace{1.5cm}
\begin{tabular}{cc}
\epsfig{file=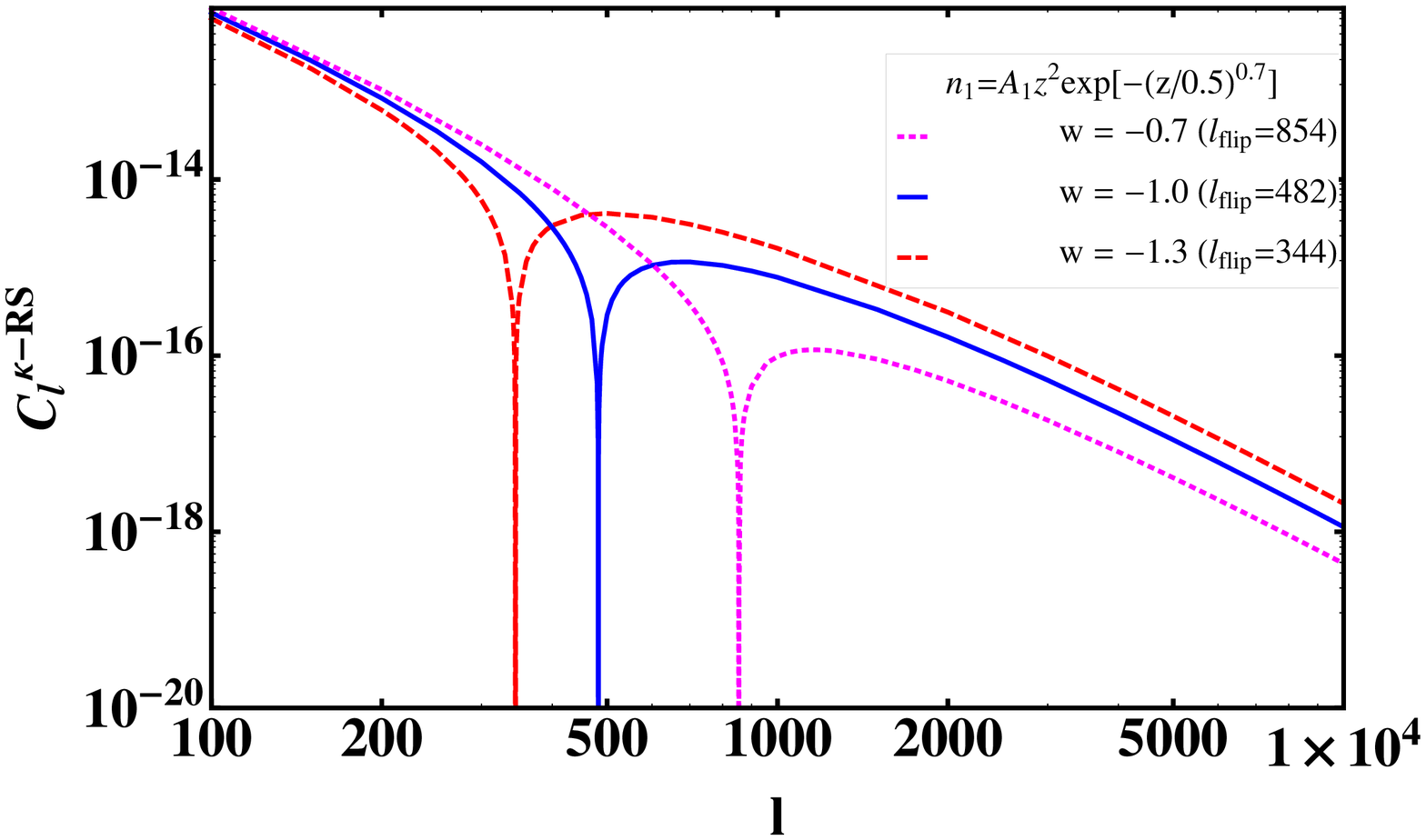,width=0.53\linewidth,clip=} &
\epsfig{file=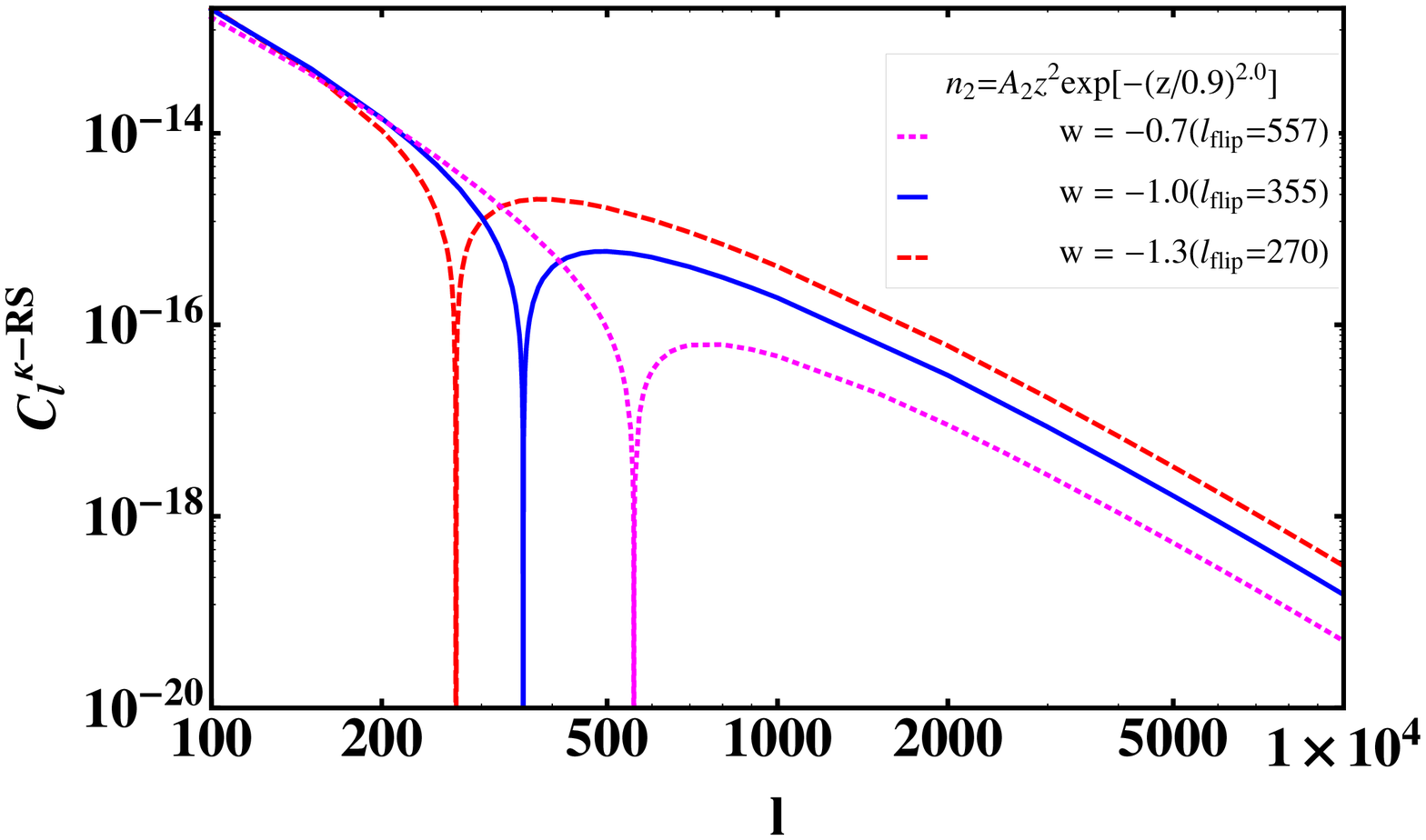,width=0.53\linewidth,clip=} \\
\end{tabular}
\vspace{-0.5cm}
\caption{CMB-WL cross-correlation spectra, $C_{l}^{\rm{RS}-\kappa}$ of different dark energy models for different WL surveys ($n_{1(2)}$). a) $C_{l}^{\rm{RS}-\kappa}$ for $n_1$ (deep survey). The dotted, the solid, and the dashed lines represent $\omega = -0.7, -1.0$, and $-1.3$, respectively.  b) $C_{l}^{\rm{RS}-\kappa}$ for $n_2$ (shallow survey) with the same notation as the left panel. }. \label{fig1}
\end{figure}

We show the CMB-WL cross-correlation, $C_{l}^{\rm{RS}-\kappa}$ of different dark energy models for two WL surveys in Fig.\ref{fig1}. This figure depicts the absolute values of $C_{l}^{\rm{RS}-\kappa}$. In the left panel of Fig.\ref{fig1}, we use the deep WL survey ($n_1 = 0.64 z^2 \exp [-(z/0.5)^{0.7}]$) to investigate the dark energy dependence of $C_{l}^{\rm{RS}-\kappa}$. The dashed, the solid, and the dotted lines correspond to $\omega = -1.3, -1.0$, -0.7, respectively. $C_{l}^{\rm{RS}-\kappa}$ is anti-correlated at large angles and changes to the positively-correlation at small angles. These flipping scales for different dark energy models $\omega = -1.3, -1.0$, and -0.7 are $l_{\rm{flip}} = 344, 482$, and 854, respectively. In the right panel of Fig.\ref{fig1}, the dark energy dependence of $C_{l}^{\rm{RS}-\kappa}$ is shown for the shallow WL survey ($n_1 = 3.10 z^2 \exp [-(z/0.9)^2]$). Flipping scales for different dark energy models $\omega = -1.3, -1.0$, and -0.7 are $l_{\rm{flip}} = 270, 355$, and 557, respectively. Thus, the deep survey is prominent for the investigation of dark energy using the flipping scale. Also, the deep survey yields the better signal to noise ratio compared to the shallow one \cite{07111696}.    

\begin{figure}
\centering
\vspace{1.5cm}
\begin{tabular}{cc}
\epsfig{file=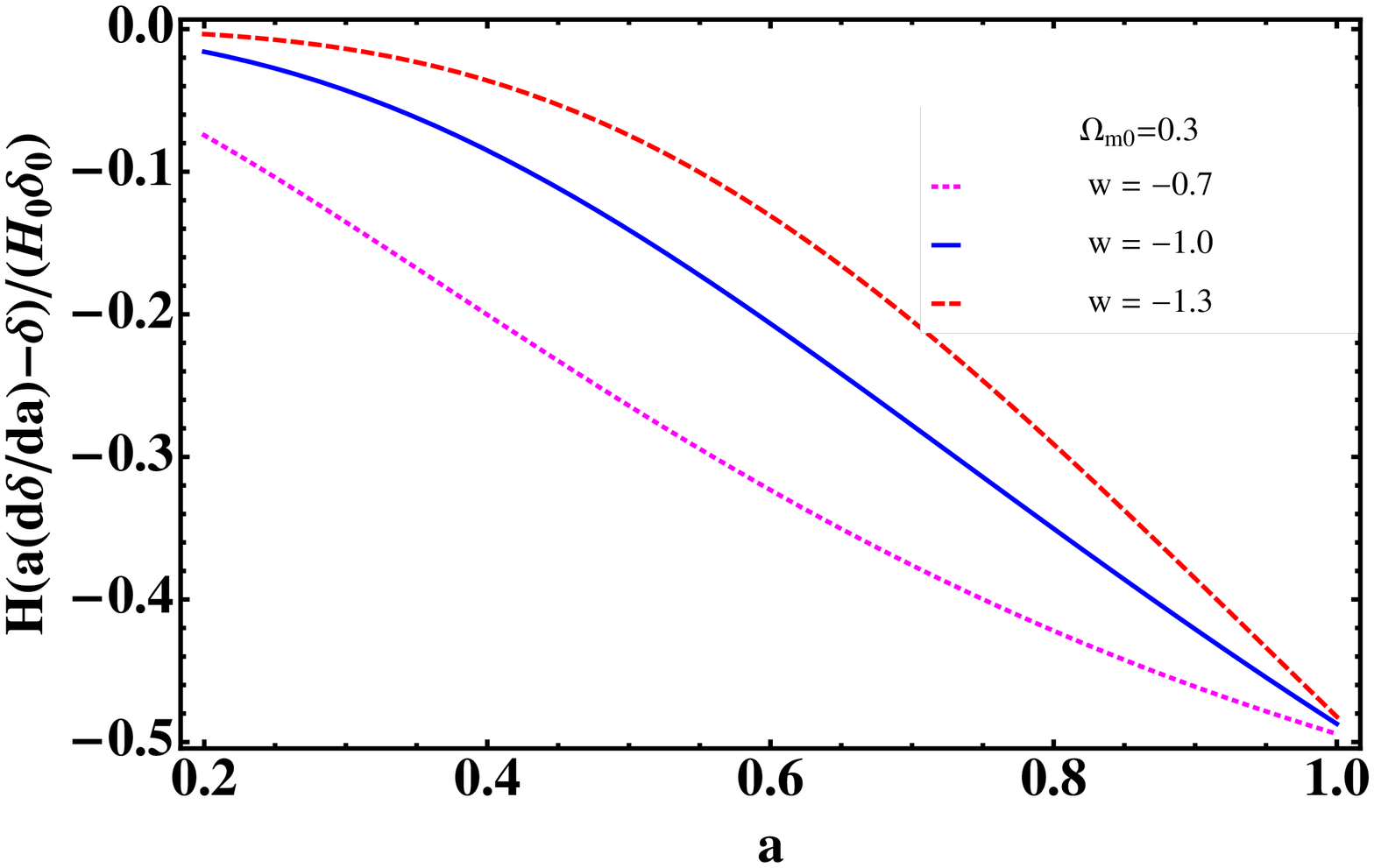,width=0.53\linewidth,clip=} &
\epsfig{file=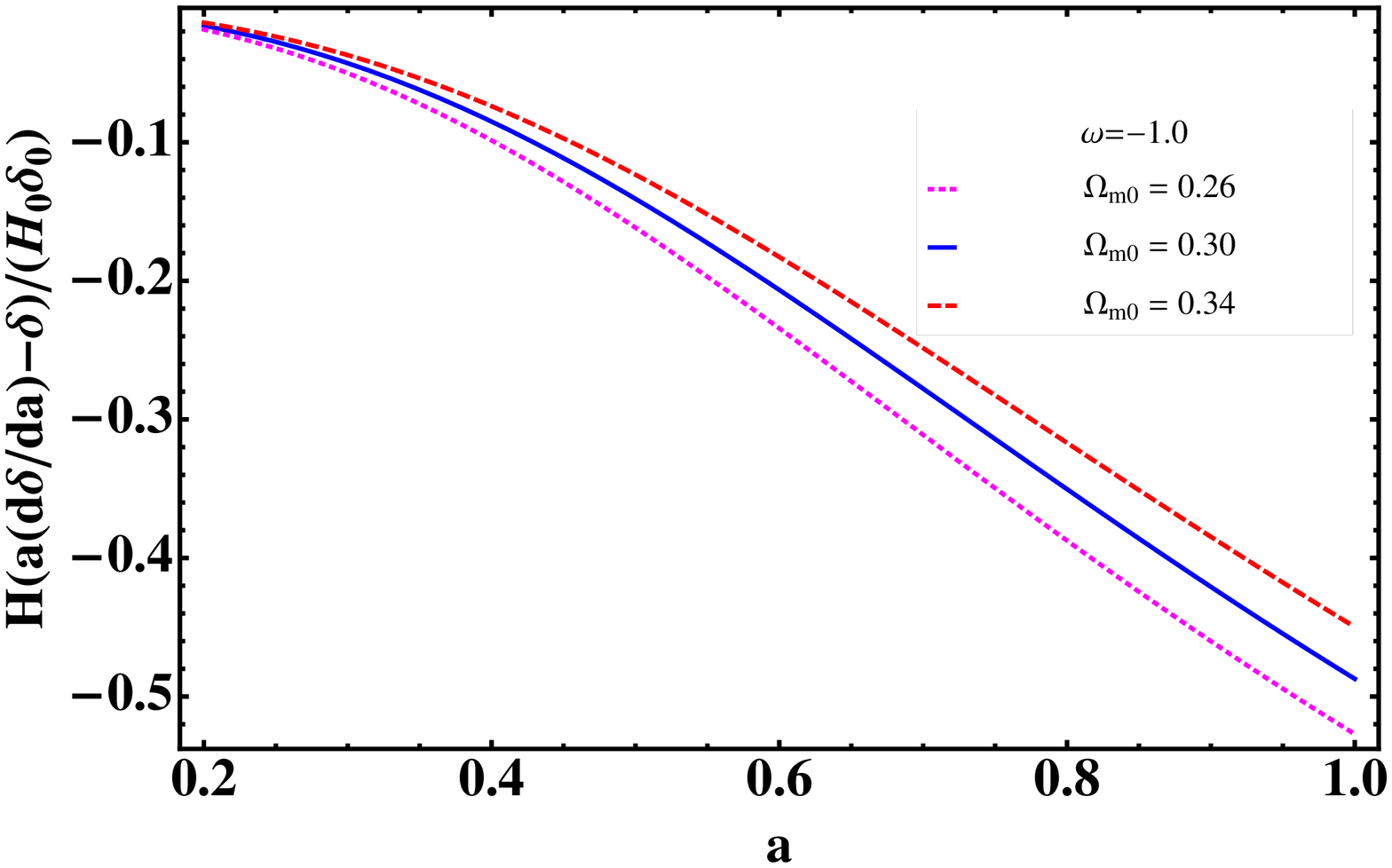,width=0.53\linewidth,clip=} \\
\end{tabular}
\vspace{-0.5cm}
\caption{Evolutions of $(\delta' - a H \delta)/a$ for different $\omega$ and $\Omega_{\rm{m}0}$ in the linear regime. a) Evolutions of $(\delta' - a H \delta)/a$ for different DE models when $\Omega_{\rm{m}0} = 0.3$. The dotted, the solid, and the dashed lines represent $\omega = -0.7, -1.0$, and $-1.3$, respectively.  b) Evolutions of $(\delta' - a H \delta)/a$ of the $\Lambda$CDM model for different values of $\Omega_{\rm{m}0}$. The dotted, the solid, and the dashed lines correspond to $\Omega_{\rm{m}0} = 0.26, 0.30$, and $0.34$, respectively.  }. \label{fig2}
\end{figure}

One can understand the DE dependence of the flipping scale from the Eq. (\ref{Phip}). The sign change of the cross-correlation $P_{\Phi \Phi'}$ is caused by the sign change of $\Phi'$. For the overdensity, $\Phi$ is always negative in both linear and non-linear regimes. However, $\Phi'$ is positive for the overdensity in the linear regime but $\Phi'$ becomes negative in some non-linear scales. This scale is matched when $\delta' - a H \delta = 0$. One can rewrite this equation as 
\be \delta' - a H \delta = a H \delta \Bigl( \fr{d \ln \delta}{d \ln a} -1 \Bigr) = \fr{d \ln (\delta/a)}{d \eta} \times \delta  \label{deltapm} \, . \ee
Thus, if the overdensity grows faster than the background scale factor, then $\delta' - a H \delta$ becomes positive. This situation happens only in the non-linear regime. Thus, the flipping scale dependence on the dark energy corresponds to the scale dependence of $\delta/a$. We already showed behaviors of the linear regime $\delta' - a H \delta$ for different DE models in our previous work \cite{150703725}. The smaller the $\omega$, the larger the $\delta' - a H \delta$. Thus, the smaller $\omega$ DE model requires the less contribution from the non-linear to convert the sign of $\Phi'$. That is the reason, the smaller value of flipping scale, $l_{\rm{flip}}$ for the smaller value of $\omega$. The dependence of the location of $l_{\rm{flip}}$ on DE is degenerated with the changing the value of $\Omega_{\rm{m}0}$. The larger the $\Omega_{\rm{m}0}$, the larger the $\delta' - a H \delta$. Thus, the flipping scale, $l_{\rm{flip}}$ becomes larger as one increases $\Omega_{\rm{m}0}$. This is the reason why there exists the difference between the value for $l_{\rm{flip}}$ of our $\Lambda$CDM model and one ($l_{\rm{flip}} \sim 800$) in reference \cite{07111696}. In this reference, the adopted value for the $\Omega_{\rm{m}0}$ is 0.26. In this manuscript, we use $\Omega_{\rm{m}0} = 0.3$ and the value of $l_{\rm{flip}}$ should be smaller than that is in the reference. In order to understand these $\omega$ and $\Omega_{\rm{m}0}$ dependences of the flipping scale,  the linear evolution of $(\delta' - a H \delta)/a$ are presented in Fig.\ref{fig2}. In the left panel of Fig.\ref{fig2}, the evolutions of $(\delta' - a H \delta)/a$ for different DE models are depicted when $\Omega_{\rm{m}0} = 0.3$. The dashed, solid, and dotted lines correspond to $\omega = -1.3, -1.0$, and -0.7, respectively. As we explain before, $(\delta' - a H \delta)/a$ is lager for the smaller value of $\omega$. Thus, the flipping scale value becomes smaller as $\omega$ decreases. In the right panel of Fig.\ref{fig2}, we show the evolutions of $(\delta' - a H \delta)/a$ of $\Lambda$CDM model for different values of $\Omega_{\rm{m}0}$. The dashed, solid, and dotted lines represent $\Omega_{\rm{m}0} = 0.34, 0.30$, and 0.26, respectively. As $\Omega_{\rm{m}0}$ increases, the linear $(\delta' - a H \delta)/a$ gets close to zero. Thus, the flipping scale becomes smaller as $\Omega_{\rm{m}0}$ increases. This shows the degeneracy between $\omega$ and $\Omega_{\rm{m}0}$ on $l_{\rm{flip}}$. The degeneracy tendencies between $\omega$ and $\Omega_{\rm{m}0}$ are different at different redshifts. However, the cross-correlation of RS-WL is redshift independent and one might not be able to remove degeneracy  between $\omega$ and $\Omega_{\rm{m}0}$ on the location of the flipping scale. Thus, if one can use the redshift dependent observables on the flipping scale, then it will be more powerful for the investigation of the DE than the RS-WL correlation.  

One might be able to use the cross-correlation between the momentum field and the density field instead of RS-WL. Because the momentum is defined as $\vec{p} = (1+\delta) \vec{v}$ and the momentum divergence is equal to $\delta' = - \nabla \cdot \vec{p} \equiv - q$ from the continuity equation. Using the momentum field also has the advantage because the momentum is zero in voids. One can obtain the power spectrum of momentum field estimated directly from the observed radial velocity data at galaxy position \cite{0012066}. Thus, one might obtain $P_{\delta\delta'}(k,z) - aH P_{\delta\delta}(k,z)$ at different redshifts from the cross-correlation of the divergence of the momentum field and the density field and the power spectrum of the density field itself. And this will be the more powerful method than RS-WL cross-correlation. 

\section{Conclusions}
We show that the flipping scale of the cross-correlation between the CMB RS effect and the WL convergence dependence on the dark energy. As the dark energy equation of state, $\omega$ increases, the angular value of flipping scale, $l_{\rm{flip}}$ decreases. Thus, one might use this flipping scale as the investigation of the dark energy. Unfortunately, this scale dependence on the $\omega$ is degenerated with $\Omega_{\rm{m}0}$. However, if one can use the redshift dependence observables as the flipping scale, then one might be able to break this degeneracy because the they have the different tendencies. One might be able to use the cross-correlation between the divergence of the momentum field and the density field for this purpose. This is under the progress.   

\section*{Acknowledgments}
This work were carried out using computing resources of KIAS Center for Advanced Computation.

\end{document}